\documentclass[onecolumn,showpacs,preprintnumbers,amsmath,amssymb]{revtex4}

\usepackage{bm}

\def \in #1 #2 {\int \limits_{#1}^{#2}}

\usepackage{amssymb}
\usepackage{amsmath}
\usepackage{graphicx} 
\usepackage{color}

\usepackage{amsmath}

\usepackage{pstricks}

\def\localinput#1{{
  \renewcommand{\documentclass}[2][dummy]{}
  \renewcommand{\usepackage}[2][dummy]{}
  \renewenvironment{document}{}{}
  \def\jobname{#1}
  \input{#1}
}}

\def\sla#1{\ooalign{\hfil\hspace{-0.1ex}\raise.2ex\hbox{$\not \phantom{#1}$}\hfil\crcr  $#1$}}

\begin{document}

\preprint{}

\title{A New Proposal for Neutrino Mass and $|V_{ud}|$ Measurements}

\author{Akihiro Matsuzaki}
 \email{akihiro@rikkyo.ac.jp }
\affiliation{%
Department of Physics, Rikkyo University,
Nishi-ikebukuro, Toshima-ku Tokyo, Japan, 171
}%

\author{Hidekazu Tanaka}%
 \email{tanakah@rikkyo.ac.jp}
\affiliation{%
Department of Physics, Rikkyo University,
Nishi-ikebukuro, Toshima-ku Tokyo, Japan, 171
}%

\date{\today}

\begin{abstract}
    We introduce a new method to detect the absolute neutrino mass scale.
    It uses a macroscopic mass of tritium source.
    We explain that the neutrino mass can be measured by scaling the mass difference of the source between initial and final state, and its heat value.    
    This method is free from the electron energy resolution limit and the statistical error.
    We estimate the required accuracy to measure the neutrino mass.
    We also report that the $\{u,d\}$ component of the CKM matrix, $|V_{ud}|$ may be determined in $10^{-6}$ accuracy as an application of this work.
\end{abstract}

\pacs{Valid PACS appear here}
\maketitle


\section{Introduction}
    The neutrino was introduced by E. Fermi in 1935 \cite{Fermi}. 
    Since then, the neutrino is treated as the massless fermion.
    However, the neutrino oscillation demands at least two of three neutrinos to have non-zero masses, and the squared mass splittings are \cite{PDG}
\begin{align} \begin{split}
\Delta m_{12}^2&=(6-9)\times 10^{-5}\ [\mathrm{eV}^2],\\
\Delta m_{23}^2&=(1-3)\times 10^{-3}\ [\mathrm{eV}^2].
\end{split} \end{align}    
   
    Meanwhile, the neutrino mass has been searched by many experimentists. 
    Today, one of the most established method is as follows:
    Prepare the tritium (T) sample.
    Tritium decays into helium-3 ion, electron, and anti-electron-neutrino.
    This process is written as 
\begin{align} \begin{split}
\mathrm{T}\to\mathrm{{}^3He^+} + e^- + \bar \nu_e.
\end{split} \end{align}    
    Then detect the electron energy distribution.
    The endpoint spectrum of this distribution depends on the neutrino mass.
    So, you can determine the neutrino mass by detecting the endpoint, accurately. 
    The latest upper bound is $m_\nu<2$ eV \cite{PDG}.
    Some other methods are performed, for example, the neutrinoless double beta decay and the cosmological observation, and they give the upper bounds $\langle m_\nu\rangle < 0.7 - 2.8$ eV \cite{NEMO3} and $\Sigma m_\nu< 2.0$ eV \cite{WMAP}, respectively.

    If the mass hierarchy is normal, the lower bound of second heavy neutrino mass is $m_{2}>0.008$ eV as the lightest neutrino is massless.
    If we have this sensitivity, we surely determine the absolute neutrino mass.

    However, the KATRIN experiment, which will start in 2012 \cite{KATRIN}, is designed with sensitivity to measure the effective neutrino mass $m_\nu>0.2$ eV.

    The absolute neutrino mass scale is one of the undetermined parameter of the Standard Model (SM).
    Some models beyond the SM, for example, GUTs \cite{GUTs1, GUTs2} predict the absolute neutrino mass scale.
    Moreover, for reliability, the alternative method to detect the neutrino mass is important.  

    Our new method explained in this paper is one of the tritium beta decay experiments. 
    However, today's tritium beta decay experiments are suffered from some difficulties as follows:
    First, the required electron energy resolution is more and more severe.
    Next, the produced electron energy spectrum is distorted by random multiple scattering in the source.
    These difficulties are caused by detecting the electron kinetic energy.

    The KATRIN experiment \cite{KATRIN} uses the huge detector which spectrometer is 23 m long and the gaseous tritium source will consist of a 10 m long.
    Our method needs less space since we measure the mass difference and heat value of macroscopic sample as explained later.

    This paper is organized as follows:
    In Section \ref{s2}, we explain the new method.
    In Section \ref{s3}, we derive the required accuracy to measure the neutrino mass.
    In Section \ref{s4}. we discuss the results and summarize this work.

\section{The New Method}\label{s2}

    The concept of the new method we introduce is as follows:
    Elementally, the beta decay process is written as $n^0\to p^+ + e^- + \bar \nu_e$.
    In this process, as explained later, 
     the $\bar \nu_e$ mean energy $\langle E_\nu \rangle$ depends on the effective $\bar \nu_e$ mass $m_\nu$, which is explained in Appendix \ref{Aeff}. 
    Then, we can determine $m_\nu$ by detecting $\langle E_\nu \rangle$.

    This method has some advantages compared to the existing ones.
    In final state of beta decay, $\bar \nu_e$ has no interaction with the sample.
    Since $\langle E_\nu \rangle$ is the mean value, we don't have to measure the beta decays event-by-event.
    Then, we can enlarge the mass of the sample to the macroscopic scale.

    We employ the tritium source because the half-life of tritium is approximately 12.33 years;
    energy difference between the initial and final state particles is small;
    tritium is easy to obtain;
    tritium has a small nucleon number.
    The long lifetime enables us to detect the mass difference and the heat value precisely.
    The small energy difference suppresses the increasing temperature 
    (as explained in Appendix \ref{app3}), 
    and all of the produced electrons are captured in the sample. 
    Small nucleon number corresponds to the large events par unit mass.


    To determine $\langle E_\nu \rangle$ experimentally, we introduce $\Sigma E_\nu$ as the total energy of produced neutrinos during the experiment, and $N$ as the number of beta decay events during it.
    Using them, we have $\langle E_\nu \rangle=\Sigma E_\nu/N$.
    Since we cannot detect $\Sigma E_\nu$ directly, it must be treated as a missing energy.
    Then, we consider the energy conservation between the initial and final state sources.
    In the initial state, we set the condensed tritium which have the mass $M_i$ in a container.
    It includes some impurities.
    The mass of the container and the impurities is represented as $\chi$.
    In the final state, some of tritium decay into ${}^3\mathrm{He}^+ +e^- +\bar \nu_e$.
    $\bar \nu_e$ have no interaction and go away.
    ${}^3\mathrm{He}^+ $ and $e^-$ have too small energy to penetrate the container (typically they have some keV momenta), then they stay in the container and construct the ${}^3\mathrm{He}$ atoms.
    The sum of ${}^3\mathrm{He}$ produced in the experiment and the remained tritium have mass $M_f$.
    The sample releases thermal energy $M_Q$ since the beta decay is an exothermal reaction.
    Assuming that the sample is isothermal during the experiment, $M_Q$ is dissipated away from the sample by the heat conduction and radiation.
    Therefore, the total mass in the final state is $M_f+\chi$.
    Then, the energy conservation between the initial and final states gives
\begin{align} \begin{split}
M_i+\chi=M_f+\chi+M_Q+\Sigma E_\nu.
\end{split} \end{align}
    On the other hand, the number of decay event $N$ is given by 
\begin{align} \begin{split}\label{n}
N=\frac{\Delta M}{m_{\mathrm{T}}-m_{\mathrm{He}}},
\end{split} \end{align}
     where $\Delta M \equiv M_i-M_f$, and $m_{\mathrm{T}}$ and $m_{\mathrm{He}}$ are the masses of tritium and helium-3 atoms, respectively.
    Using these equations, the mean neutrino energy is given by
\begin{align} \begin{split}\label{a}
\langle E_\nu \rangle&=\frac{\Sigma E_\nu}{N} 
= (m_{\mathrm{T}}-m_{\mathrm{He}})\left(1-\frac{M_Q }{\Delta M }\right).
\end{split} \end{align}

    Therefore, we have to detect $m_{\mathrm{T}}-m_{\mathrm{He}}$, $M_Q $, and $\Delta M $ to determine $\langle E_\nu \rangle$.
    First, $m_{\mathrm{T}}-m_{\mathrm{He}}$ should be given by other experiments as explained in Appendix \ref{appQ}.
    Second, to detect $M_Q $, we introduce the technology of differential scanning calorimeter \cite{ThermochimicaActa397155}. 
\begin{figure}
  \begin{center}
    \includegraphics[keepaspectratio=true,height=50mm]{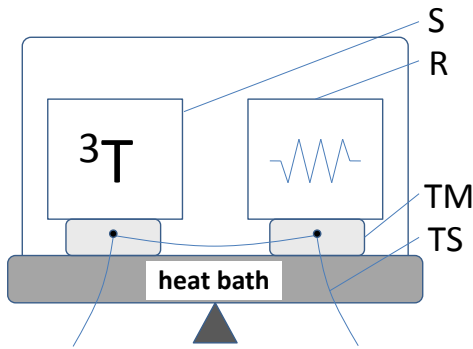}
  \end{center}
  \caption{The basic concept of the new method.
            We detect the mass deference and the heat value.}
  \label{fig:fig1.eps}
\end{figure}
    Fig. \ref{fig:fig1.eps} is a schematic drawing of the new method.
    S is a test sample and R is a reference one.
    S and R are located symmetrically.
    R releases the thermal energy controlled by the electric heater.
    TM is the thermo-module and TS is the thermo sensor which measures the difference of released thermal energy between S and R.
    Measuring the electric energy which is injected to cancel the temperature difference between S and R,
we can determine the $M_Q$ as released thermal energy from S. 
    Last, $\Delta M $ is detected by the weighting machine.



    Now we study the neutrino mass dependence of $\langle E_\nu \rangle$.
    The leading order differential decay width which contains exact $m_\nu$ effect takes the form  
\begin{widetext}\begin{align} \begin{split}\label{dG/dE}
\frac{d\Gamma}{dE_\nu}
&=\frac{G_F^2|V_{ud}|^2 }{(2\pi)^3}
\frac{\sqrt{(E_\nu^2-m_\nu^2)\bigl\{(M_3-2m_{\mathrm{T}} E_\nu)^2-4m_e^2(M_4-2m_{\mathrm{T}} E_\nu)\bigr\}}}{M_4-2m_{\mathrm{T}} E_\nu}
\\&\times
\biggl[
 \Bigl\{
  (1+C_A)^2E_\nu(M_1-2m_{\mathrm{T}} E_\nu)+(1-C_A^2)m_{\mathrm{He}^+}(M_3-2m_{\mathrm{T}} E_\nu)
 \Bigr\}
\\& +\frac{1}{2}\Bigl\{(1-C_A)^2M_2-2(1-C_A^2)m_{\mathrm{He}^+} m_{\mathrm{T}}\Bigr\}
\frac{(M_3-2m_{\mathrm{T}} E_\nu)(m_{\mathrm{T}}-E_\nu)}{M_4-2m_{\mathrm{T}} E_\nu}
\\& -(1-C_A)^2m_{\mathrm{T}}
\Bigl\{\frac{3(M_3-2m_{\mathrm{T}} E_\nu)^2(m_{\mathrm{T}}-E_\nu)^2}
{6(M_4-2m_{\mathrm{T}} E_\nu)^2}
+\frac{(E_\nu^2-m_\nu^2)\bigl\{(M_3-2m_{\mathrm{T}} E_\nu)^2-4m_e^2(M_4-2m_{\mathrm{T}} E_\nu)\bigr\}}
{6(M_4-2m_{\mathrm{T}} E_\nu)^2}
\Bigr\}
\biggr],
\end{split} \end{align}\end{widetext}
    where $G_F$, $V_{ud}$, $C_A$, $m_e$, and $m_{\mathrm{He}^+}$ are the Fermi constant, the $\{u, d\}$ component of the CKM-Matrix, the axial current coupling constant, electron mass, and helium-3 ion mass, respectively; 
    and
\begin{align} \begin{split}
\\ M_1&\equiv  m_{\mathrm{T}}^2-m_e^2+m_\nu^2-m_{\mathrm{He}^+}^2
\\ M_2&\equiv  m_{\mathrm{T}}^2+m_e^2-m_\nu^2-m_{\mathrm{He}^+}^2
\\ M_3&\equiv  m_{\mathrm{T}}^2+m_e^2+m_\nu^2-m_{\mathrm{He}^+}^2
\\ M_4&\equiv  m_{\mathrm{T}}^2+m_\nu^2.
\end{split} \end{align}

    The neutrino mean energy is given by
\begin{align} \begin{split}\label{aen}
\langle E_\nu \rangle \equiv
\frac{\int dE_\nu E_\nu \frac{d\Gamma}{dE_\nu}}{\int dE_\nu \frac{d\Gamma}{dE_\nu}},
\end{split} \end{align}
    where
\begin{align} \begin{split}\label{int}
\int dE_\nu &\equiv \int_{m_\nu}^{E_{\nu}^{ \mathrm{Max}}}  dE_\nu,
\end{split} \end{align}
    and
\begin{align} \begin{split}
E_{\nu}^{ \mathrm{Max}}&={\frac{M_1-2 m_e m_{\mathrm{He}^+} }{2 m_{\mathrm{T}} }}.    
\end{split} \end{align}

\section{The Required Accuracy}\label{s3}

    $\langle E_\nu\rangle$ is affected not only by $m_\nu$ but also by $C_A$, $m_\nu$, $m_{\mathrm{He}^+}$, $m_e$, and $m_{\mathrm{T}}-m_{\mathrm{He}^+}-m_e$. 
    How accurate must we measure these quantities to detect the neutrino mass?
    To estimate this, we introduce
\begin{widetext}
\begin{align} \begin{split}
\delta\langle E_\nu(m_\nu; \delta C_A,\delta m_{\mathrm{He^+}},\delta m_e ) \rangle 
\equiv
  \frac{\langle E_\nu(m_\nu; C_A+\delta C_A,m_{\mathrm{He^+}}+\delta m_{\mathrm{He^+}},m_e+\delta m_e )\rangle-\langle E_\nu(0; C_A,m_{\mathrm{He^+}},m_e)\rangle}
       {\langle E_\nu(m_\nu; C_A+\delta C_A,m_{\mathrm{He^+}}+\delta m_{\mathrm{He^+}},m_e+\delta m_e )\rangle+\langle E_\nu(0; C_A,m_{\mathrm{He^+}},m_e)\rangle}
.
\end{split} \end{align}\end{widetext}
    Here we use the quantities, $C_A=12.6$, $m_{\mathrm{He^+}}=2800$ MeV, $m_e=0.511$ eV, $m_\mathrm{T}-m_{\mathrm{He^+}}-m_e=0.2$ MeV, and
    $\delta C_A,\delta m_{\mathrm{He^+}},\delta m_e $ mean the errors of $ C_A, m_{\mathrm{He^+}}, m_e $, respectively.

    Actually, $C_A$, $m_{\mathrm{He^+}}$, $m_e$, and $m_\mathrm{T}-m_{\mathrm{He^+}}-m_e$ given above are not the true values, and also the Eq. (\ref{dG/dE}) does not give the accurate decay width since it contains no corrections.
    However, it does not matter for our purpose.
    We just want to estimate the order of $m_\nu$ effect and errors caused by $\delta C_A$, $\delta m_{\mathrm{He^+}}$, and so on.

    We first consider $\langle E_\nu \rangle$ differences caused by $m_\nu$ as 
\begin{align} \begin{split}
 \delta\langle E_\nu( 0.1\ \mathrm{eV};0,0,0)\rangle
   &\simeq 1.12\times10^{-11},
\\ \delta\langle E_\nu( 1\ \mathrm{eV};0,0,0)\rangle
 &\simeq 1.12\times10^{-9}.
\end{split} \end{align}
    The required accuracy on $\langle E_\nu\rangle$ increase by two digits as the neutrino mass which we want to measure becomes one-tenth.

    On the other hand, the normalized error caused by $C_A$ is estimated as
\begin{align} \begin{split}
 \delta\langle E_\nu( 0;10^{-4},0,0)\rangle
  &\simeq 2.44\times10^{-11},
\\ \delta\langle E_\nu( 0;10^{-2},0,0)\rangle
  &\simeq 2.43\times10^{-9}.
\end{split} \end{align}
    The required accuracy on $C_A$ increases by one digit as the required accuracy on $\langle E_\nu\rangle$ increases by one digit.
    The reason why the required accuracy on $C_A$ is milder than that on $\langle E_\nu\rangle$ is explained in Appendix \ref{appCa}.

    After all, we have to measure $C_A$ in $10^{-4}$ ($10^{-2}$) order to detect $0.1$ eV (1 eV) neutrino mass 
    because to detect the neutrino mass excess, the $C_A$ originated error in $\langle E_\nu\rangle$ has to be less than the $\langle E_\nu\rangle$ variance caused by $m_\nu$.

\subsection{Other Uncertainties }

    To measure $m_\nu$ accurately, we must overcome some other uncertainties.
    They are namely $m_{\mathrm{He^+}}$, $m_e$, $m_{\mathrm{T}}-m_{\mathrm{He}^+}-m_e$, $M_i-(M_f+M_Q)-\Sigma E_\gamma$, and $M_Q+\Sigma E_\gamma$.

    First, we consider $\delta\langle E_\nu(0;0,\delta m_{\mathrm{He}^+},0)\rangle$. 
    Fixing other variables, it is given by     
\begin{align} \begin{split}
\delta\langle E_\nu(0;0,2800\times10^{-8}\ \mathrm{MeV},0)\rangle
   &\simeq -1.54\times10^{-14},
\\  \delta\langle E_\nu(0;0,2800\times10^{-10}\ \mathrm{MeV},0)\rangle
  &\simeq -1.54\times10^{-16}.
\end{split} \end{align}

    Next, we consider $\delta\langle E_\nu(0;0,0,\delta m_e)\rangle$. 
    Fixing other variables, it is given by       
\begin{align} \begin{split}
\delta\langle E_\nu(0;0,0,0.511\times 10^{-8}\ \mathrm{MeV})\rangle
   &\simeq -1.45\times10^{-11},
\\\delta\langle E_\nu(0;0,0,0.511\times 10^{-10}\ \mathrm{MeV})\rangle
   &\simeq -1.45\times10^{-13}.
\end{split} \end{align}

    Third, we consider $\langle E_\nu\rangle' \equiv \langle E_\nu(m_{\mathrm{T}}-m_{\mathrm{He}^+}-m_e)\rangle /(m_{\mathrm{T}}-m_{\mathrm{He}^+}-m_e)$ error caused by $m_{\mathrm{T}}-m_{\mathrm{He}^+}-m_e$ uncertainty.
    Fixing other variables as before, it is given by       
\begin{align} \begin{split}
  \frac{\langle E_\nu(0.02\times(1+10^{-8})\ \mathrm{MeV})\rangle'-\langle E_\nu(0.02\ \mathrm{MeV})\rangle'}
       {\langle E_\nu(0.02\times(1+10^{-8})\ \mathrm{MeV})\rangle'+\langle E_\nu(0.02\ \mathrm{MeV})\rangle'}
  &
\simeq 1.46\times10^{-11},
\\\frac{\langle E_\nu(0.02\times(1+10^{-10})\ \mathrm{MeV})\rangle'-\langle E_\nu(0.02\ \mathrm{MeV})\rangle'}
       {\langle E_\nu(0.02\times(1+10^{-10})\ \mathrm{MeV})\rangle'+\langle E_\nu(0.02\ \mathrm{MeV})\rangle'}
 &
\simeq 1.46\times10^{-13}.
\end{split} \end{align}
    It is reasonable to consider $\langle E_\nu\rangle'$ because in Eq. (\ref{a}), $m_{\mathrm{T}}-m_{\mathrm{He}}$ varies along with $\langle E_\nu (m_{\mathrm{T}}-m_{\mathrm{He}^+}-m_e)\rangle$, and $m_{\mathrm{T}}-m_{\mathrm{He}}$ is easy to calculate from $m_{\mathrm{T}}-m_{\mathrm{He}^+}-m_e$.
    
    These errors are proportional to $\langle E_\nu\rangle$ error.
    If we want to measure $m_\nu$ to 0.2 eV as the KATRIN goal, $\langle E_\nu\rangle$ has to be determined with $\delta \langle E_\nu\rangle<10^{-11}$, and then $\delta m_{\mathrm{He}^+}/m_{\mathrm{He}^+}<10^{-5}$, $\delta m_e/m_e<10^{-8}$, and $\delta(m_{\mathrm{T}}-m_{\mathrm{He}^+}-m_e)/(m_{\mathrm{T}}-m_{\mathrm{He}^+}-m_e)<10^{-8}$, respectively.

    Last, $\Delta M$ and $M_Q$ in Eq. (\ref{a}) require the same accuracy as $\langle E_\nu\rangle$.

    As a result, the relations between $m_\nu$ which we want to detect and the required accuracies of related variables are shown in Table \ref{table1}. 

\begin{widetext}
\begin{center}
\begin{table}[h]
 \caption{
 The relation between $m_\nu$ which we want to detect and the required accuracies of related variables.
            For example, if $m_\nu=0.2$ eV, then we have to determine $C_A$ as $\delta C_A/C_A<10^{-4}$, also $m_{\mathrm{He^+}}$ as $\delta m_{\mathrm{He^+}}/m_{\mathrm{He^+}}<10^{-5}$, etc.
          }
\label{table1}
\begin{center}
  \begin{tabular}{|c||l|l|l|l|l|
  }
\hline
 $m_\nu$ we want to detect
                  & 20 eV       &  2 eV       &  0.2 eV      &  0.02 eV     &  0.002 eV  
\\
\hline
\hline
 $\langle E_\nu\rangle$
                  &  $10^{-7}$  & $10^{-9}$   & $10^{-11}$   & $10^{-13}$   & $10^{-15}$  
 \\
\hline
\hline
 $C_A$            &  $1$        & $10^{-2}$   & $10^{-4}$    &  $10^{-6}$   &  $10^{-8}$ 
\\
\hline
 $m_{\mathrm{He}^+}$       &  $10^{-1}$  & $10^{-3}$   & $10^{-5}$    & $10^{-7}$    & $10^{-9}$  
\\
\hline
 $m_e$,\ \ \ 
$m_\mathrm{T}-m_{\mathrm{He}^+}-m_e$&  $10^{-4}$  & $10^{-6}$   & $10^{-8}$    & $10^{-10}$   & $10^{-12}$ 
\\
\hline
$\Delta M$,\ \ \ $M_Q$ &  $10^{-7}$  & $10^{-9}$   & $10^{-11}$   &  $10^{-13}$  &  $10^{-15}$ 
\\
\hline
  \end{tabular}
 \end{center}
\end{table}
\end{center}
\end{widetext}

    The required accuracy in $C_A$ is comparatively low.
    This is because $C_A$ in Eq. (\ref{aen}) does not appear in the leading order of $E_\nu^{\mathrm{Max}}$. 
    This situation is true for other corrections, for example, radiative correction and nucleus-dependent correction.
    The errors in the Fermi constant and CKM-matrix do not affect $\langle E_\nu \rangle$ at all.
    The Fermi function for the Coulomb correction is calculable.

\section{$|V_{ud}| $ measurments}

    As an application of this work, we can determine $\Gamma$ (lifetime), $|V_{ud}|$, and $C_A$ accurately up to theoretical uncertainty. 


    The number of decayed tritium $N_1$ and $N_2$ at time $t_1$ and $t_2$, respectively, are given by
\begin{align} \begin{split}
N_1&=N_0(1-e^{-\Gamma t_1})\\
N_2&=N_0(1-e^{-\Gamma t_2}),
\end{split} \end{align}
    where $N_0$ is the number of tritium in the initial state.
    Then, defining the ratio $N_2/N_1\equiv A$, $\Gamma$ and $A$ are related as 
\begin{align} \begin{split}
\frac{d A}{A}&=\left( \frac{\Gamma t_1 e^{-\Gamma t_1}}{1-e^{-\Gamma t_1}}-\frac{\Gamma t_2 e^{-\Gamma t_2}}{1-e^{-\Gamma t_2}} \right) \frac{d \Gamma}{\Gamma}\\
&\simeq \frac{\Gamma }{2}(t_2-t_1)\frac{d\Gamma}{\Gamma},
\end{split} \end{align}
    where we use the approximations $\Gamma t_1\ll 1$ and $\Gamma t_2\ll 1$ in the second line.
    Therefore, if we measure $d A/A=\mathcal{O}(10^{-7})$ for a two-year experiment, the $\Gamma$ ambiguity is $d\Gamma / \Gamma =\mathcal{O}(10^{-6})$ since $\Gamma(t_2-t_1)/2\sim 10^{-1}$. 
    We note here that $A$ is actually determined without $m_\mathrm{T}-m_{\mathrm{He}}$ value since $A$ can be written as
\begin{align} \begin{split}\label{eq A}
A \equiv \frac{N_2}{N_1}=\frac{\frac{M_i-M_{f2}}{m_\mathrm{T}-m_{\mathrm{He}}}}{\frac{M_i-M_{f1}}{m_\mathrm{T}-m_{\mathrm{He}}}}
= \frac{\Delta M_2}{\Delta M_1},
\end{split} \end{align}
    where the indices $1$ and $2$ represent the values at time $t_1$ and $t_2$, respectively.


    $\Gamma$s contain $|V_{ud}|$ and $C_A$ as parameters.
  If we can determine $C_A$ with $10^{-6}$ accuracy, which is for example realized by the $10^{12}$ of ${}^3\mathrm{He}^{2+} +e^- \to \mathrm{T}^++\nu_e$ scattering events or determining the decay width of another nuclear species with $10^{-6}$ accuracy,
    then, we can determine $|V_{ud}|$ with $10^{-6}$ accuracy.
    Alternatively, if we have the ability to detect the $0.02$ eV $m_\nu$, we can determine $C_A$ with $10^{-6}$ accuracy.
    Then we can determine $|V_{ud}|$ with $10^{-6}$ accuracy.


    According to Ref. \cite{PDG}, today's experimental value is $|V_{ud}|=0.97418 \pm 0.00027$.
    Hence, this method may have a great impact not only on the lepton sector but also on the quark sector.

\section{Summary and Discussion}\label{s4}


    We explained the new method to detect the neutrino mass.
    This method asks the precision measurement of mass defect and the released thermal energy.
    The results are written up in TABLE \ref{table1}.

    We also pointed out that this method is useful for $|V_{ud}|$ precision measurement.
    This is because we can determine $\Gamma$ precisely in this method.
    
    In TABLE \ref{table1}, the error of $m_\mathrm{T}-m_{\mathrm{He}^+}-m_e$ will reduce to $10^{-6}$ in near future as explained in Appendix \ref{appQ}.
    Also, the error of $M_Q$ can be already reduced to $10^{-8}$ by the present technology (Appendix \ref{app3}).
    
    In Eq. (\ref{n}) and then also in Eqs. (\ref{a}) and (\ref{eq A}), we dealt with $M_i-M_f$.
    It is modified by considering the heat capacity of the sample.
    However, its effect can be controlled easily if the temperature of the sample is kept in low as explained in Appendix \ref{Ap Heat Capacity}.    


    Today, the experiments of elementary particle physics are categorized as the accelerator or non-accelerator ones.
    In both of them, the detected quantities are energy and momentum of the particle, which are the microscopic quantities.
    However, our method is an experiment which reveals the properties of elementary particles using the macroscopic quantities.

%
%
The normalized standard deviation of $\langle E_\nu\rangle$ in one beta-decay event is $\sigma/\langle E_\nu\rangle=0.303$ and for 1 mol tritium beta decays, it becomes
\begin{align} \begin{split}
\frac{\sigma/\sqrt{1[\mathrm{mol}] \times N_A}}{\langle E_\nu\rangle}
\simeq3.91\times10^{-13},
\end{split} \end{align}
    where $N_A=6.02\times 10^{23}[\mathrm{mol}^{-1}]$ is the Avogadro constant.
    Therefore, we do not have to consider seriously the statistical error for $0.2$ eV neutrino mass. 
    Our work is unique in this respect.

\section*{Acknowledgement}
    We would like to thank associate professor Jiro Murata for his valuable advice.


\appendix

\section{Effective Neutrino Mass}\label{Aeff}
    The decay width written in the mass eigenstates $m_k$ is approximated as 
\begin{align} \begin{split}
&\Gamma=\sum_{k=1}^3 |U_{ek}|^2 \Gamma(m_k)
\simeq \sum_{k=1}^3 |U_{ek}|^2\Gamma(0)+\sum_{k=1}^3 |U_{ek}|^2 \frac{m_k^2}{2}\frac{d^2\Gamma(m_k)}{dm_k^2}|_{m_k=0}
\\&=\Gamma(0)+ \frac{m_\nu^2}{2}\frac{d^2\Gamma(m_\nu)}{dm_\nu^2}|_{m_\nu=0}
\simeq \Gamma(m_\nu),
\end{split} \end{align}
    where $U_{ek}$ are the leptonic mixing matrix elements, and
\begin{align} \begin{split}
   \sum_{k=1}^3 |U_{ek}|^2   =  1   ,\hspace{5em}
 \sum_{k=1}^3 |U_{ek}|^2 m_k^2 \equiv  m_\nu^2.
\end{split} \end{align}   
    As you can see from Eqs. (\ref{dG/dE}) and (\ref{int}), the linear term does not appear.

\section{Heat Value of the Source}\label{app3}
    When a T decays into ${}^3$He${}^++e^-+\bar \nu_e$, the electron and ${}^3\mathrm{He}^+$ have about 6 keV momentum in total.
    This becomes the thermal energy in source.
    If we suppose that the pure tritium source has 3 g mass, it has about $N_0=$6.02 $\times10^{23}$ tritium atoms.
    According to the tritium half-life (12.33 years), the heat value in one second is 
\begin{align} \begin{split}
& N_0 (1-e^{-\Gamma t})\bigr|_{t=1[\mathrm{sec}]}\times 6 [\mathrm{keV}]
\simeq 1[\mathrm{J}].
\end{split} \end{align}

    Ref. \cite{ThermochimicaActa397155} explains that the differential scanning calorimetry has the 25 [nW] resolution. 
    This is $\mathcal{O}(10^{-8})$ of 1 [J/s].

\section{Suppression of $C_A$ Dependence in $\langle E_\nu \rangle$ }\label{appCa}
    Here, we show how the $C_A$ dependence in $\langle E_\nu \rangle$ is suppressed.
    For simplicity, we represent the differential decay width as
\begin{align} \begin{split}
\frac{d\Gamma}{dE_\nu}&\sim
 (a E_\nu^2+b\frac{E_\nu^3}{m_{\mathrm{T}}})
+C_A(cE_\nu^2+d\frac{E_\nu^3}{m_{\mathrm{T}}}),  
\end{split} \end{align}
    where the coefficients $a$, $b$, $c$, and $d$ are in the same order.
    Hence, the integral is easy to evaluate: 
\begin{align} \begin{split}
\int dE_\nu \frac{d\Gamma}{dE_\nu}&\sim
 (\frac{a}{3} {E_{\nu}^\mathrm{ Max}}^3+\frac{b}{4}\frac{{E_{\nu}^\mathrm{ Max}}^4}{m_{\mathrm{T}}})
+C_A(\frac{c}{3}{E_{\nu}^\mathrm{ Max}}^3+\frac{d}{4}\frac{{E_{\nu}^\mathrm{ Max}}^4}{m_{\mathrm{T}}}),  
\\\int dE_\nu E_\nu \frac{d\Gamma}{dE_\nu}&\sim
 (\frac{a}{4} {E_{\nu}^\mathrm{ Max}}^4+\frac{b}{5}\frac{{E_{\nu}^\mathrm{ Max}}^5}{m_{\mathrm{T}}})
+C_A(\frac{c}{4}{E_{\nu}^\mathrm{ Max}}^4+\frac{d}{5}\frac{{E_{\nu}^\mathrm{ Max}}^5}{m_{\mathrm{T}}}).  
\end{split} \end{align}
    Then, for ${E_{\nu}^\mathrm{ Max}}/m_{\mathrm{T}}\ll 1$, the mean neutrino energy becomes
\begin{align} \begin{split}
\langle E_\nu \rangle=\frac{\int dE_\nu E_\nu \frac{d\Gamma}{dE_\nu}}{\int dE_\nu \frac{d\Gamma}{dE_\nu}}
\sim
\frac{3}{4}{E_{\nu}^\mathrm{ Max}}
( 1 +\frac{b+d C_A}{a+c C_A}\frac{1}{20}\frac{{E_{\nu}^\mathrm{ Max}}}{m_{\mathrm{T}}}).
\end{split} \end{align}
    In tritium beta decay, the suppression factor ${E_{\nu}^\mathrm{ Max}}/(20m_{\mathrm{T}})$ is $\mathcal{O}( 10^{-6})$.

\section{$m_\mathrm{T}-m_{\mathrm{He}^+}-m_e$}\label{appQ}

    Ref. \cite{0904.4712} says that "In the history of mass spectrometry the precision of atomic mass determination has shown a constant improvement of about an order of magnitude every decade".
    Also ref. \cite{PRL70.2888} reported that the mass deference between tritium and helium-3 is 18590.1(1.7) eV in 1993.
    These facts suggest that we will be able to determine $m_\mathrm{T}-m_{\mathrm{He}^+}-m_e$ with $10^{-6}$ error in 2013.
    Actually, the Max Planck Institute for Nuclear Physics aims for $ 10^{-7}$ according to Ref. \cite{PhysReports425.1}.

\section{Heat Capacity}\label{Ap Heat Capacity}


    If we consider the heat capacity of the sample, the equation (\ref{n}) is replaced by 
\begin{align} \begin{split}\label{n2app}
N=\frac{M_i'-M_f'}{m_{\mathrm{T}}-m_{\mathrm{He}}},
\end{split} \end{align}
    where $M_i'$ and $M_f'$ are the masses of initial and final state sample which are defined at zero temperature, respectively.
    These have the relations, $M_i = M_i'+ C_i T$ and $M_f = M_f'+ C_i T$, where $T$ is the sample temperature, and $C_i$ and $C_f$ are the sample heat capacity in the initial and final state, respectively.
    Then, the mean neutrino energy is given by
\begin{align} \begin{split}\label{aapp}
\langle E_\nu \rangle&
=(m_{\mathrm{T}}-m_{\mathrm{He}})\left(1-\frac{M_Q-T \Delta C }{\Delta M-T \Delta C }\right)\\
&\simeq (m_{\mathrm{T}}-m_{\mathrm{He}})\left(1-\frac{M_Q }{\Delta M }\right)\left(1+\frac{T\Delta C }{\Delta M }\right),
\end{split} \end{align}
    where $\Delta C \equiv C_i-C_f$.
    In the second line, we have used an approximation, $T \Delta C/\Delta M\ll1$.


    Also, the equation (\ref{eq A}) becomes 
\begin{align} \begin{split}
A =\frac{M_i'-M_{f2}'}{M_i'-M_{f1}'}
\simeq \frac{\Delta M_2}{\Delta M_1}\left(1+\frac{T \Delta C_1}{\Delta M_1}-\frac{T \Delta C_2}{\Delta M_2}\right),
\end{split} \end{align}
    where the indices $1$ and $2$ represent the values at time $t_1$ and $t_2$, respectively. 
    
    We can estimate the magnitude of these correction terms according to the Debye model.
    The Debye temperature $T_D$ of $\mathrm{H}_2$ is 105 K.
    Then, in low temperature ($T\ll T_D$), the specific heat is expressed as
\begin{align} \begin{split}
C_V&\simeq 234\times N_A k_B \left(\frac{T}{T_D}\right)^3
\simeq 234\times (6.02\times 10^{23}\ [\mathrm{mol}^{-1}]) \times (1.38\times 10^{-23}\ [\mathrm{J/K}])\left(\frac{10}{105}\right)^3 \left(\frac{T}{10 \mathrm{K}}\right)^3 \\
&=0.168 \times \left(\frac{T}{10 \mathrm{K}}\right)^3  [\mathrm{J/(K\cdot mol)}],
\end{split} \end{align}
    where $k_B=1.38\times 10^{-23}\ [\mathrm{J/K}]$ is the Boltzmann constant and $N_A=6.02\times 10^{23}\ [\mathrm{mol}^{-1}]$ is the Avogadro constant.
    5.5 \% of tritium decay in one year since the half-life of tritium is about 12.32 year.
    Assuming that tritium and helium-3 have the same order of specific heat, we give $\Delta C\sim 0.168\ [\mathrm{J/(K\cdot mol)}]\times 1\ [\mathrm{mol}] \times 0.055\simeq 0.01$ [J/K].
    Also, $\Delta M \simeq 18 \mathrm{keV} \times N_A \times 0.055\times (1.6\times 10^{-19}\mathrm{[J/eV]})\simeq 10^8$ J.
    Then $T \Delta C/\Delta M \sim 10 [\mathrm{K}] \times 0.01[\mathrm{J/K}] / 10^8[\mathrm{J}] \simeq 10^{-9} $.
    This means that we can easily estimate the heat capacity effect enough accurately in the low temperature system.

\end{document}